\documentstyle[12pt,epsfig]{article}
\def\mm{ {\mathbf{m}}}
\def\kk{ {\mathbf{k}}}

\def\noi{ \noindent }

\def\beq{\begin{equation}}
\def\enq{\end{equation}}

\def\beqa{\begin{eqnarray}}
\def\eeqa{\end{eqnarray}}

\begin{document}

\title{ Metric Features of a Dipolar Model}
\author{
Mario Casartelli$^{1,2,\dag}$, Luca Dall'Asta $^{3,\ddag}$, Enrico
Rastelli $^{2,4}$ and Sofia Regina $^{2}$ \\
}

\setcounter{page}{0}
\maketitle \thispagestyle{empty}

\begin{abstract}
\noindent The lattice spin model, with nearest neighbor
ferromagnetic exchange and  long range dipolar  interaction, is studied by the method
of time series for observables based on cluster configurations and
associated partitions, such as Shannon entropy, Hamming and Rohlin
distances. Previous results based on the two peaks shape of the
specific heat, suggested the existence of two possible
transitions. By the analysis of the Shannon entropy we are able to
prove that the first one is a true phase transition corresponding
to a particular melting process of oriented domains, where colored
noise is present almost independently of true fractality. The
second one is not a real transition and it may be ascribed to a
smooth balancing between two geometrical effects: a progressive
fragmentation of the big clusters (possibly creating fractals),
and the slow onset of a  small clusters chaotic phase. Comparison
with the nearest neighbor Ising ferromagnetic system points out a
substantial difference in the cluster geometrical
properties of the two models and in their critical behavior.
\end{abstract}

\vspace{2cm}
\noi $^1$ Dipartimento di Fisica dell'Universit\`a - Parco Area Scienze 7a,
43100 PR (Italy)\\

\noi $^2$ CNR - Istituto Nazionale di Fisica della Materia, Parma 43100 PR (Italy)\\

\noi $^3$ Laboratoire de Physique Th\'eorique, Universit\'e de Paris-Sud, 91405
ORSAY (France) \\

\noi $^4$ CNR - IMEM (Parma)\\

\noi $^\dag$ Electronic address: casartelli@fis.unipr.it\\
\noi $^\ddag$ Electronic address: luca.dallasta@th.u-psud.fr\\

\newpage
\noindent{\bf 1. Introduction}

There is a growing literature illustrating the conceptual and
practical relevance of two dimensional (2D) systems with long
range interactions (see \cite{booth} \cite{ifti} and references
therein). The model we consider here is an Ising model on a square
lattice, with both nearest neighbor (NN) ferromagnetic exchange
interaction, and long range dipolar interactions decaying as
$~r^{-3}~$ among  all pairs in the lattice ($~r$ being the
distance). Spins are supposed to be perpendicular to the lattice
plane. We shall denote PFD such a perpendicular (P) ferromagnetic
(F) dipolar (D) system. This model shows a very interesting
thermodynamic behavior that results, at low temperature, in the
presence of regularly shaped stripes of upwards and downward
spins, and, at increasing temperature, in a complex onset of
disorder  until the usual random paramagnetic phase occurs. It has
been supposed that one or more transitions could take place in the
region between the ordered and paramagnetic phases. In particular,
Ifti and coworkers \cite{ifti} studied by numerical Monte Carlo
(MC) simulations the behavior of specific heat, obtaining a curve
with a sharp peak at temperature $T_1$ and a broad maximum in the
region at $T_2 \sim 2T_1$. The authors suggested  the existence of
two possible phase transitions, the former related to the melting
of the stripes, the latter to the occurrence of the paramagnetic
phase. We shall study this item by MC simulation introducing a new
method of investigation that seems to disprove this conjecture, in
favor of a single phase transition.

Since the pioneering works by Peierls and Griffith \cite{PG}, the
shape and distribution of the magnetic clusters have been
suggested to be a significant geometrical signature of 2D models.
The problem is to give quantitative estimates and qualitative
connections, beside visual inspection, between the cluster
features and the thermodynamic behavior of the system. To this
end, we shall  extend to  PFD an analysis already tested in other
contexts, such as microcanonical and canonical Ising models, or
self organized criticality (see
\cite{parti1}\cite{parti2}\cite{soc}). The basic tool is a map
between the space $  \mathcal{C}\equiv \mathcal{C (\mathbf{M})} $
of configurations on the lattice $\mathbf{M}$ and a ``partition
space" $ \mathcal{Z}\equiv \mathcal{Z(\mathbf{M})}$, defined by
the correspondence between homogeneous connected clusters and
subsets of the lattice. When a dynamical simulation is performed
on $\mathbf{M}$, we look for possible meaningful relations between
geometrical and dynamical features of quantities in $\mathcal{C}$
and $\mathcal{Z}$ and the physically relevant (thermodynamic)
properties. This may be done by a time series analysis of
observables related to the metric properties of $\mathcal{C}$ and
$\mathcal{Z}$. The method is very general, and its efficiency
consists precisely in giving indications not exclusively tailored
on the model, making possible  comparisons with other systems and
other dynamics.

 The Shannon entropy, for instance, points
out the order-disorder transition by a sudden change of its slope
as a function of temperature. Since the entropy continuously
depends on the cluster measure distribution, this transition may
be read as a topological breakdown driven by channels joining the
stripped domains of the ground state. On the contrary, the
order-disorder transition in the NN Ising model (studied in
\cite{parti2}) is driven by a fractal fragmentation of the
clusters, leading to a sharp increase of entropy. Standard
deviations, in both models, develop a singularity. It seems
therefore that the analysis of this quantity can give information
about the kind of incoming disorder. Further information can be
obtained from time series of distances in $
\mathcal{C(\mathbf{M})}$ and $\mathcal{Z(\mathbf{M})} $, from
their standard deviation and from the analysis of power spectra,
showing the dependence of ``color exponents'' on temperature.

 In addition, for PFD, a careful examination of
clusters proves to be useful in recovering, along new and more
efficient lines, previously introduced parameters and criteria
\cite{booth} \cite{ifti}. Notations, definitions and elementary
properties of $ \mathcal{C(\mathbf{M})}$ and
$\mathcal{Z(\mathbf{M})} $, as well as Shannon entropy, Hamming
and Rohlin distances, are recalled in the Appendix, with some
mathematical details.

\vskip 30.0 pt
 \noindent{ \bf 2. The Model}

 The Hamiltonian of the 2D perpendicular Ising ferromagnet
 with dipolar interactions (PFD model), is:
 \begin{equation}\label{hamilton}
  {\mathcal{H}}= -{J\sum_{<\kk,\mm>}} s_{\kk}~ s_{\mm} +
  {g \sum_{\mm \neq \kk }  \frac{s_{\kk}~ s_{\mm}}{r_{\kk
  \mm}^{3}}}~,
\end{equation}
where $s_{\kk}$ is the usual spin variable assuming values $\pm 1$
in the lattice $\mathbf{M}$ of size $N=L\times L$. The first sum
is restricted to NN pairs, while the second sum is over all pairs.
The distance $r$ is between all sites, taking into account also
sites of periodically iterated copies of the lattice, up to the
convergence of such sums \cite{binder}. The correspondence with
the lattice $\mathbf{M}$ equipped with the binary alphabet $
{\mathbf{K}} \equiv \{0,1\}$ and knots labelled by a couple of
indexes running from $1$ to $L$ is obvious. Starting from such
$\mathbf{M}$ and $\mathbf{K}$, the mathematical apparatus
described in the Appendix I can be developed. In particular, one
may introduce the configuration space $\mathcal{C} $ provided with
the Hamming distance $d_H$, and the cluster partition space
$\mathcal{Z} $ with the Rohlin distance $d_R$.

The exchange constant $J$ in (\ref{hamilton}) and the temperature
$T$ will be given in $g$ units. A well established result (see
\cite{booth}\cite{ifti}) is that, for $J > 0.854$ the ground state
is characterized by striped domains of up and down spins, with a
trivial degeneracy corresponding to their vertical or horizontal
orientation. The stripe width $h$ increases as $J$. We shall
assume the value $J=8.9$, corresponding to $h=8$ lattice spacings.

The specific heat $C_V$ , as shown in Fig.1 versus temperature $T$
for a lattice of size $L=32$, coincides with the same quantity
shown in Fig.3 of reference \cite{booth}.
\begin{figure}[htbp]
 \centering
  \epsfig{file=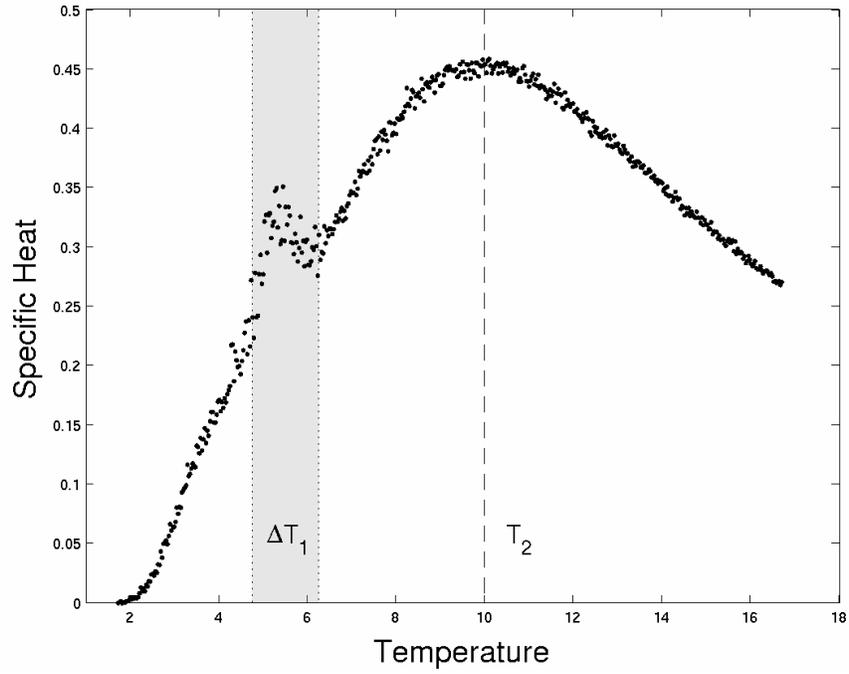,height=9cm}
  \caption{Specific Heat $C_V$, $~L=32$, mean value over two i.c..}\label{Fin1}
\end{figure}
It is characterized by two peaks: a sharp one at $T=T_1\simeq 5$,
and a broad one at $T=T_2\simeq 10$. When the lattice size is
increased,  the first peak becomes sharper and its height
increases, while the second one remains unchanged, and both are
slightly moved to lower temperatures \cite{booth}\cite{ifti}. The
temperature interval of the first peak is denoted $\Delta T_1$. In
the range $0< T < 2$, $C_V$ does not move from $0$ and stripes
remain very stable. For $2 < T < T_1$, where $C_V$ shows a sudden
rise, the jagged outline of the stripe takes place gradually. For
$T_1 < T < T_2~$ stripes are replaced by two big clusters, with
unstable appearance of small fragments (``islands''). The
relevance (in number and size) of such islands increases until the
breakdown of big clusters occurs, approaching $T_2$. Finally,
there is a progressive fragmentation into smaller and smaller
clusters. However, up to $T=16$, completely chaotic configurations
do not occur. A short summary of this process appears in Fig.2.
\begin{figure}[htbp]
  \centering
  \epsfig{file=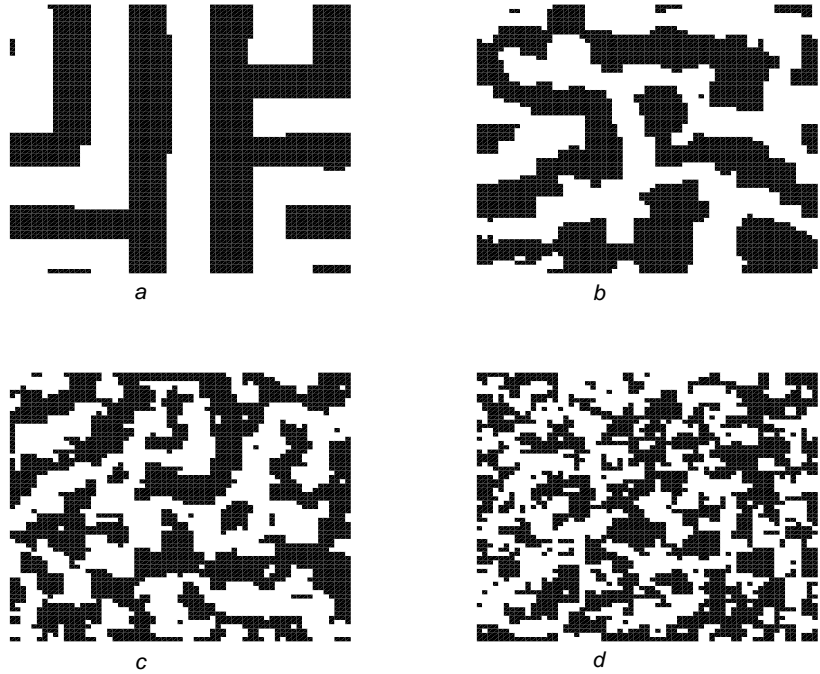,height=9cm}
  \caption{ Configurations  a, b, c, and d correspond
  to temperatures T=3, 6, 9 and 14 respectively. } \label{Fin2}
\end{figure}
In order to clarify the nature of these phases, we study the link
between geometrical and dynamical behavior in the whole range of
temperatures.

We recall that similar methods have been previously used
\cite{parti2} to investigate the Ising model (i.e. $g=0$ and $J=1$
in (\ref{hamilton})). As the phase transition is approached, at
temperature $T_c$, a sudden onset of fractal structure for the
magnetization cluster distribution occurs, with a singular
behavior of parameters like Shannon entropy or Hamming and Rohlin
distances. For instance, the standard deviation of the Shannon
entropy (see Appendix) along the orbit shows a very neat peak,
proving the onset of time instability for the configuration orbit
at $T_c$ \cite{parti2}. All this was independent of the evolution
rule (both Metropolis and deterministic dynamics were used). One
wonder whether the presence of a competitive long range
interaction in the PFD model will confirm or destroy this pattern.

The question is furtherly justified by the conjecture that in a
purely dipolar model, long range interactions do not influence the
universality class of the Ising antiferromagnet \cite{macisaac}.
Is it reasonable to extend this conjecture to the relation between
PFD model and Ising ferromagnet? The problem is not trivial since,
in such case, interactions are competitive. We shall try to answer
on the basis of geometrical considerations.

\vskip 30.0 pt

\noindent{\bf 3. Numerical Experiments}

In our Monte Carlo (MC) simulations, we shall adopt the well known
 method based on Ewald sums \cite{binder}.  This consists in considering
 a very large system which can be refolded into a smaller one with a
 renormalized coupling constant. The evolution rule is
the standard Metropolis algorithm \cite{toffoli}, where the
temperature is controlled by the flip probability.

Other general data about numerical experiments are the
following:
\begin{itemize}
\item Size: simulations have been mostly performed at $L=32$ and $64$,
with many consistency checks. Of course, larger values of $L$
would be expedient, in particular to control finite size effects.
However, not only dipolar interactions imply a sudden rise of
computing time, but there are prohibitive difficulties in handling
data at increasing $L$ for entropy and Rohlin distances (see
Appendix).
  \item Thermalization: $10^4$ MC steps are disregarded to reach equilibrium.
  The simulation at temperature $T+\delta T$ uses, as starting
  configuration, the last thermalized configuration at temperature $T$.
  \item Time average interval: $\tau=2\times 10^4 $ steps after thermalization, that ensures a good
  stabilization. Therefore, for a time
  series $X \equiv \{x_k\}, \quad k=0,...,\tau$, , the computed time average
\begin{equation}\label{timeav}
  <x>_{\tau} = \frac{1}{\tau+1}\sum_{k=0}^{\tau}x_k
\end{equation}
(i.e. the usual MC thermal average) will be simply noted $<x>$, as
in the limit $\tau \to \infty$. An example of time series for
Shannon Entropy is given in Fig.3, with the histogram of
occurrences.
\begin{figure}[htbp]
 \centering
  \epsfig{file=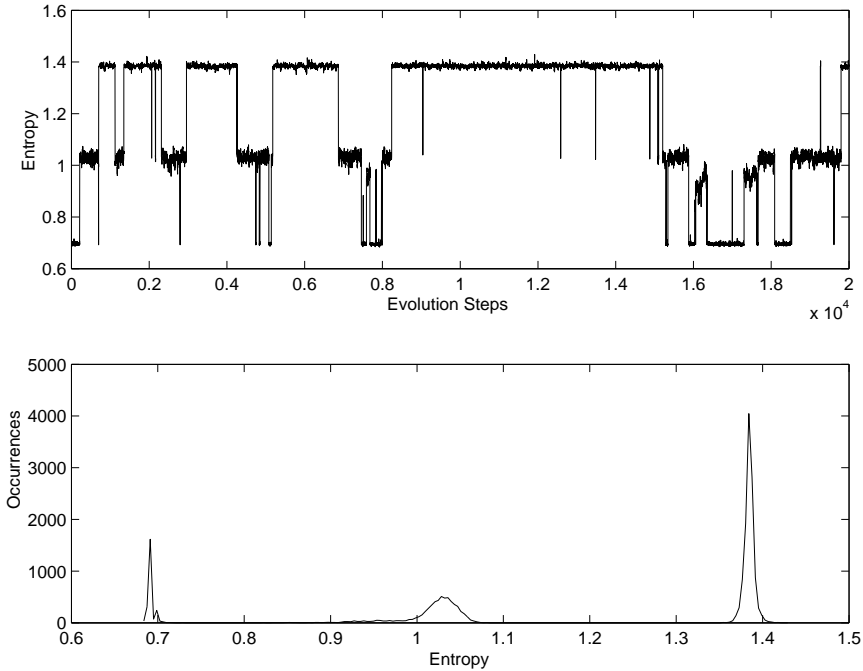,height=9cm}
  \caption{Entropy Time Series (above) and histogram of occurrences (below).
  Here $T=5.15$ and  $L=32$.}\label{Fin3}
\end{figure}

  \item Temperature range:  simulations have been performed for $2< T< 16$.
  This range has been sampled in two main ways: $100$ values with several initial conditions (i.c.), or
  $500$ values with two i.c. The two approaches  led to consistent results.
   \end{itemize}

A way to look at the meaning of time averages and their
reliability consists in looking at correlations: for a time series
$X \equiv \{x_k\}, \quad k=0,...,\tau$, where $ <x>$ is the mean
value, the self-correlation coefficient ${Corr}(X,m)$ is defined
as usual (see e.g. \cite{percival})
\begin{equation}\label{correl}
{Corr}(X,m) = \frac{\sum_{k=0}^{\tau-m}(x_{k+m}- <x> )~(x_k- <x>
)}{\sum_{k=0}^{\tau}(x_k- <x>)^2} ~.
\end{equation}
This coefficient displays how long an evolving quantity keeps the
memory of its past, measured by the lag $m$.
\begin{figure}[htbp]
  \centering
  \epsfig{file=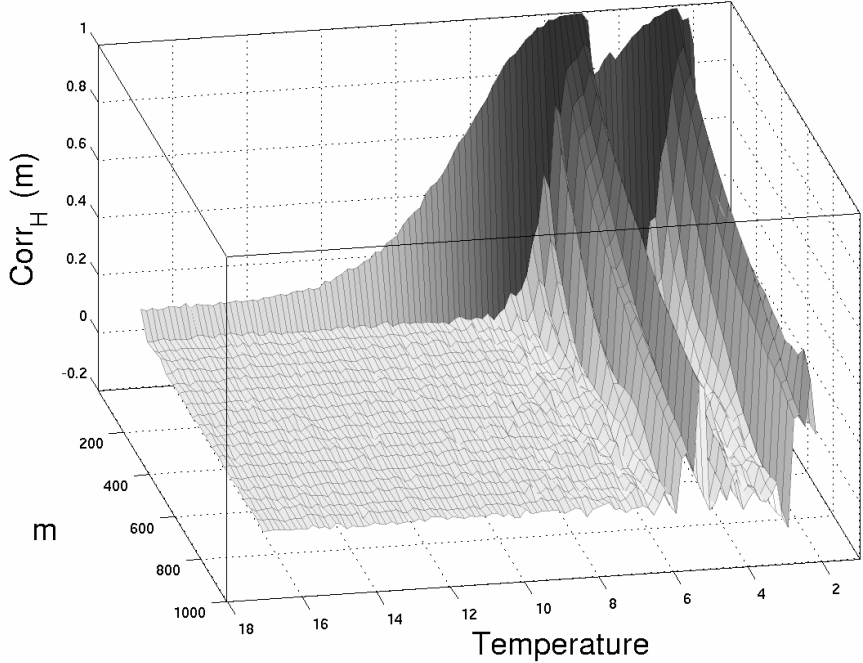,height=9cm}
  \caption{Entropy Correlation, $L=32$, mean values over four i.c.}\label{Fin4}
\end{figure}
It indicates therefore how a time (or thermal) average is built
up. For instance, in Fig.4  the  self-correlation $Corr(H,m)$ for
the entropy time series is shown (qualitatively similar results
hold for the Rohlin and Hamming distances). Fig.4 says that, apart
the trivial correlation due to the freezing at low temperature,
${Corr}(H,m)$ is sensitive to the transition at $T_1$, and rapidly
vanishes elsewhere, particularly at increasing temperature.

  Note that, below $T_1$, when the stripe boundaries
  start a certain mobility, fluctuations around the
  mean value of entropy are random for all practical purposes
  (differences between small numbers amplify the randomness of their queues):
  this is the meaning of the first valley for $2<T<T_1$.

 \vskip 30.0 pt

\noindent{\bf 4. Results}

Let's now investigate  the thermal behavior of ``geometrical''
quantities like Shannon entropy, Rohlin distance $d_R$, Hamming
distance $d_H$, their standard deviations (SD) and spectral
properties.

Shannon entropy vs. temperature is shown in Fig.5. At low
temperature, stable stripes of width $h=8$  lead correctly
   to $H=\ln4$  for $L=32$ and $H=\ln8$ for  $L=64$.
   The transition around $T_1$ is quite clear.
\begin{figure}[htbp]
  \begin{center}
  \epsfig{file=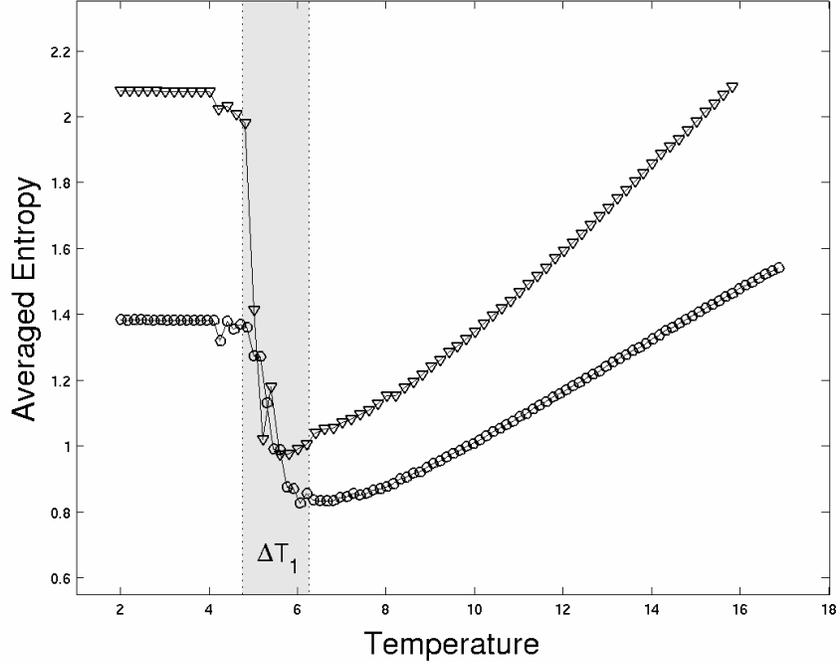,height=9cm}
  \caption{Time averaged entropy, for $L=32$ (circles) and $L=64$ (triangles),
   mean values over four i.c..}\label{Fin5}
\end{center}
\end{figure}
  More precisely, when stripes  begin to melt into two
  connected macro clusters,
  a drop is seen compatible with appearance of small spots
  (spins pointing in the opposite direction with respect to
  the background). In other terms, the Shannon entropy gives
  a quantitative evidence to the breakdown of the ground state
  connected domains, which discontinuously changes the cluster measure
  distribution by joining stripes into macro clusters. The
  example of Fig.3 clearly refers to this intermediate situation,
   when stripes still exist during long time intervals in an almost
  steady status with small oscillations at the borders, but may
  also suddenly melt or separate, modifying the cluster measures.

  The regular increase of $H$ for $T > 6$ indicates a progressive
  fragmentation of the macro clusters, or the growing relevance of islands,
  but nothing can be said about $T_2$. As to the nature of this
  fragmentation, the onset of some kind of fractality for greater $L$ cannot be excluded.
  However, for comparison, we recall that the observed fractal fragmentation
  around the transition temperature in the NN Ising ferromagnet
  gives a sudden change of concavity,
  with  vertical inflexion point, exactly at $T_c$. In conclusion, where the
  transition for PFD is confirmed  (at $T_1$), there is no fractality, and
  where fractality could be possible (around $T_2$) there is no transition:
  in both cases the difference with respect to the NN Ising model is clear.

  Also the Entropy standard deviations (or ESD), calculated for each temperature along the orbits,
  point out a critical behavior around $T_1$, followed by a regular behavior for greater $T$,
  as shown in Fig.6.
\begin{figure}[htbp]
  \centering
  \epsfig{file=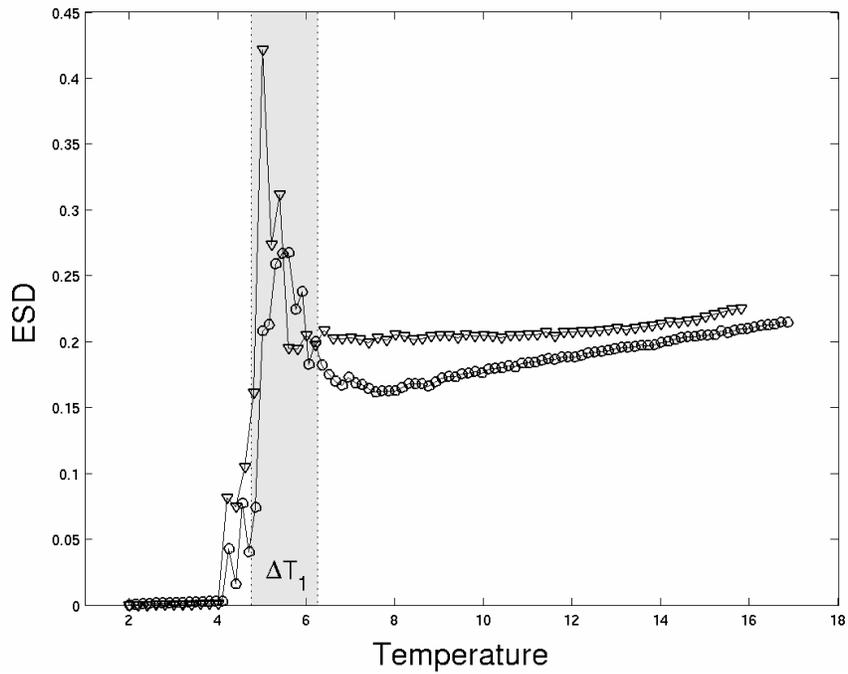,height=9cm}
  \caption{Entropy Standard Deviation in time (ESD) vs. temperature,
  same parameters and markers as in Fig.5.}\label{Fin6}
\end{figure}
   The peak at $T_1$ may be interpreted as due to time instability in the
  phase of melting stripes, corresponding indeed to intermittent behavior
  in the melting process of clusters (as illustrated by Fig.3).
  This phase is followed by the stabilization of the macro clusters
  (relative minimum of ESD). A new source of time instability is due to the cluster fragmentation,
  with appearance of islands, but once again this processes is smooth
   with respect to temperature, and no new transition can be recognized at  $T_2$.

  Fig.7 shows the Rohlin distance standard deviation
  (or RSD) versus temperature for three lattice sizes.
  \begin{figure}[htbp]
  \centering
  \epsfig{file=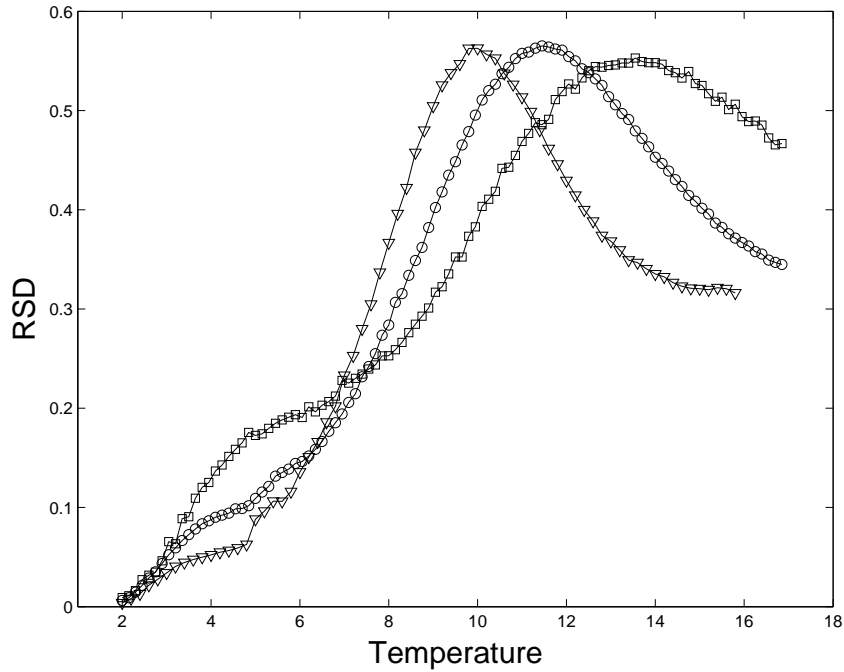,height=9cm}
  \caption{Rohlin distance Standard Deviation (RSD) in time vs. temperature,
  data for $L=16$ (squares), $L=32$ (circles)
  and $L=64$ (triangles), four i.c..}\label{Fin7}
\end{figure}
The transition at $T_1$ is not clear, even if a singularity
(discontinuity of the first derivative) seems to develop at
increasing $L$. A maximum occurs at a temperature not far from
$T_2$, where the RSD  shows an inflexion point. It is noteworthy
that the value of the maximum is independent of $L$. When compared
to the behavior of the $C_V$ peak, the behavior of this maximum
vs. $L$ is surely different. Therefore, no correlation with a new
transition can be recognized. We may see maximum of RSD in Fig.7
as the watershed between two opposite tendencies:
   1) increase of time instability, due to the rising importance
   of islands with respect to macro clusters, and
   2) the saturation of the phenomenon when macro clusters give up and the
   ensemble of disordered islands fill the
  lattice. In such a slow approach to chaoticity, the RSD
  decreases, as expected on the basis of previous experience on other models
  \cite{parti2}, where the maximum was a balancing point
  between fractal and chaotic configurations. In the present case, it would be hard to
  point out effective fractality because of small lattice sizes. As already noticed with entropy,
  a fractal phase during the fragmentation of macro clusters and the growth of islands remains
  only a reasonable conjecture, compatible with the observed behavior of RSD.

  The Hamming distance is insensitive to the cluster shape, therefore
  it is not surprising that this kind of phenomena does not appear in its standard
  deviation (HSD), as shown in Fig.8.
  On the contrary, the occurrence of a singularity  at
  $T_1$ may be sensed again for increasing $L$.
 \begin{figure}[htbp]
  \centering
  \epsfig{file=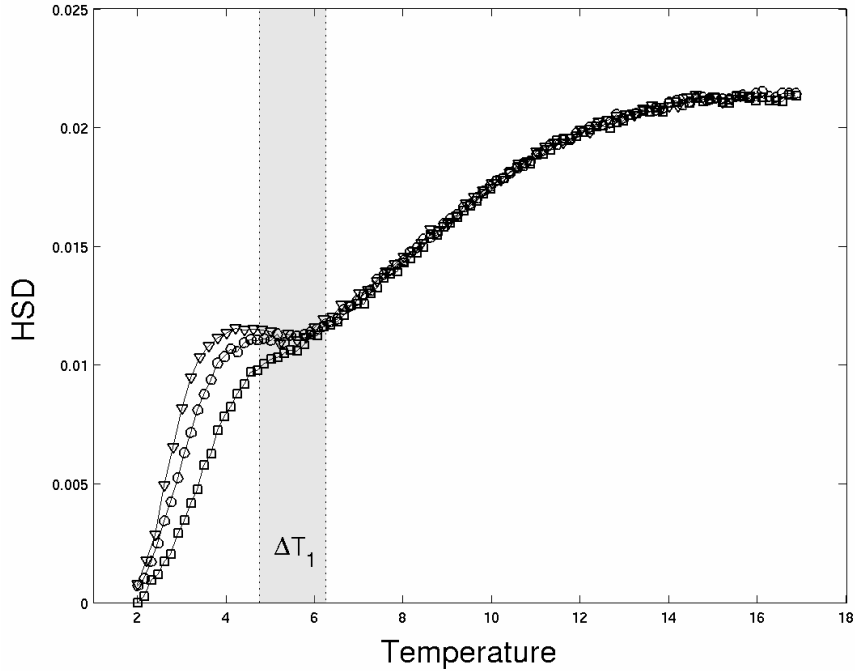,height=9cm}
  \caption{Hamming distance Standard Deviation (HSD)for PFD model, with $L=16,~32,~64$.
  The time series have   been rescaled by $L$.
  Same parameters and  markers as in Fig.7.}\label{Fin8}
\end{figure}
 Data in time series have been rescaled by
$L$, leading to a good data collapse after $T_1$. This is
consistent with the fact that, in the same range of temperatures,
mean values rescale with $N$. Moreover, such a data collapse seems
to  exclude that something may occur at $T_2$ for greater $L$.

 A comparison  with the
corresponding behavior in the Ising ferromagnet (Fig.9) is
instructive: there, the $L$-scaling behavior occurs indeed
everywhere except around the transition temperature $T_c$, where
also there is an incoming cusp that asymptotically in $L$ seems to
get an infinite derivative (in this case we reach $L=100$).
\begin{figure}[htbp]
  \centering
  \epsfig{file=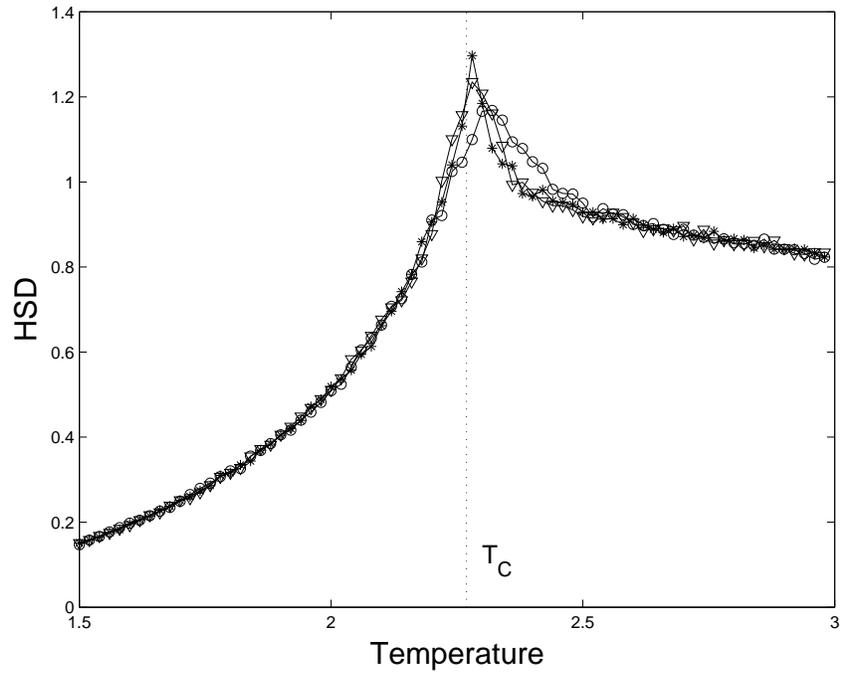,height=9cm}
  \caption{Hamming distance Standard Deviation for Ising ferromagnetic model, with
  $L=32,~64,~100$ (circles, triangles and stars respectively). Time series have been rescaled by
  $L$.}   \label{Fin9}
\end{figure}
 For both models, HSD behave qualitatively as the standard deviations of the
total cluster perimeters (we omit to report figures). Since the
perimeter SD is due to boundaries fluctuations, such a coincidence
seems to indicate a sensitivity of the Hamming distance to the
boundaries instability. A parallelism of this kind is not obvious,
considering the different role of boundaries: because of long
range interactions, indeed, the evolution rule in PFD model does
not assign to borders the same importance as in Ising model.

 Spectral features: by Fast Fourier Transform on
time series, we obtained power
  spectra ( a typical example for Entropy is shown in
  Fig.10).
\begin{figure}[htbp]
  \centering
 \epsfig{file=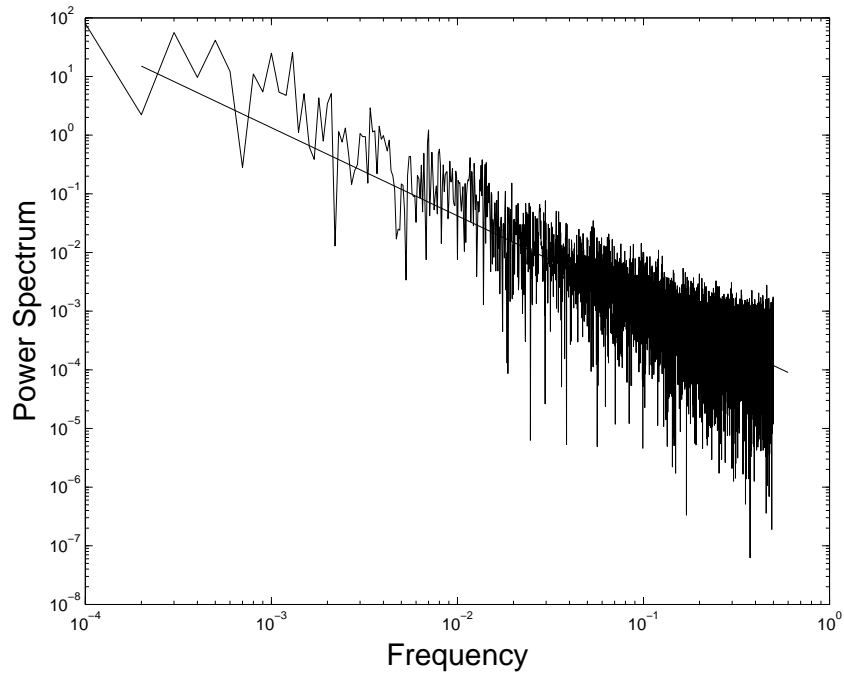,height=9cm}
  \caption{Power Spectrum and linear fit, an example on entropy time series, $T=5.4$.}\label{Fin10}
\end{figure}
   As it is well known, there is no general dynamical theory on
  the presence of colored noise in signal sources, in front of an extremely rich phenomenology
  (see e.g. \cite{percival} \cite{schroeder}). In our context, experience on comparable time series
  for other models (Ising ferromagnet and SOC)
  confirms the widely discussed empirical link between fractality and colored noise, provided
  that the lattice size is sufficient to achieve a reasonable
  fractality \cite{parti1} \cite{parti2}\cite{soc}.
  Therefore, in the present case, such a link remains
  conjectural, but this makes the observation of the noise even
  more interesting.

  As a general feature, for all observables there is in fact
  the expected tendency to chaoticity as $T$ increases, but at $T=16$
  a genuine white noise regime is not yet achieved. This agrees with the
  disordered but not completely chaotic aspect of fragmented clusters
  at the same temperature. Thus, in the whole range of interest,
  colored noise $\omega^{\alpha}~, \alpha < 0$, is the rule.
  This exponent can be obtained as the value of
  the angular coefficient from the linear fit in the loglog plot
  of the power spectrum. For instance,
   Fig.10  shows the loglog plot od the power spectrum of
   the Shannon entropy for $T=5.4$. The linear fit gives $\alpha= -1.54$.
     A simple way to get information on the dependence of the  noise
  is to plot  the exponent $\alpha$ vs. $T$.
   For entropy spectra, the result is exhibited in Fig.11.
\begin{figure}[htbp]
  \centering
  \epsfig{file=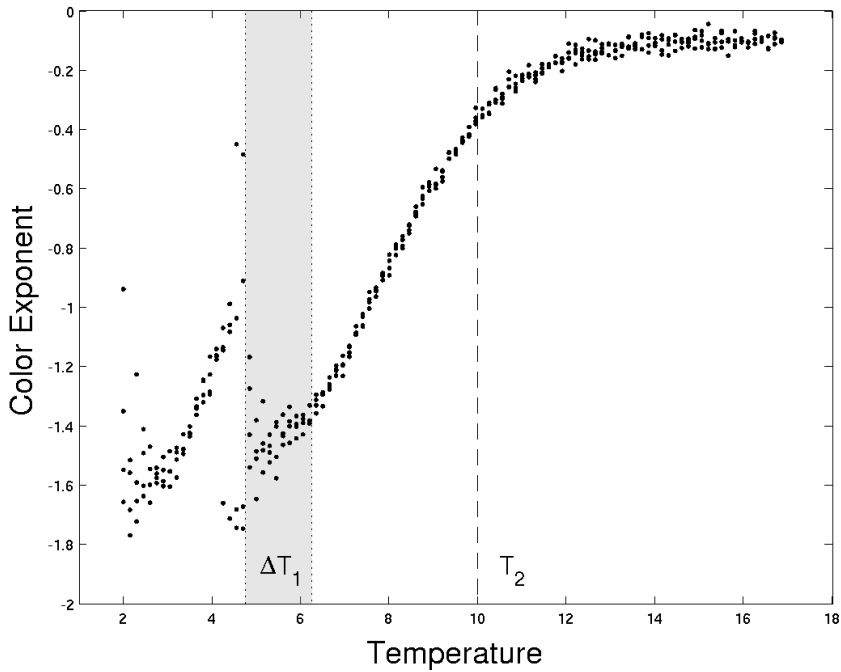,height=9cm}
  \caption{Color Exponents $\alpha$ from Entropy Spectra,
  four distinct i.c. at each temperature.}\label{Fin11}
\end{figure}
  Here we preferred not to average over different i.c., in order to stress the
  dispersion of values in the critical interval $\Delta T_1$. The maximum
  spread coincides  with the beginning of the interval
  $\Delta T_1$.  Then, after the well (coinciding with the onset of macro clusters)
  there is the expected
  slow growth, up to values close to $0$ from below. It is instructive to consider
  the analogous figure of the spectral exponent
  for the NN Ising model (Fig.12).
\begin{figure}[htbp]
  \centering
  \epsfig{file=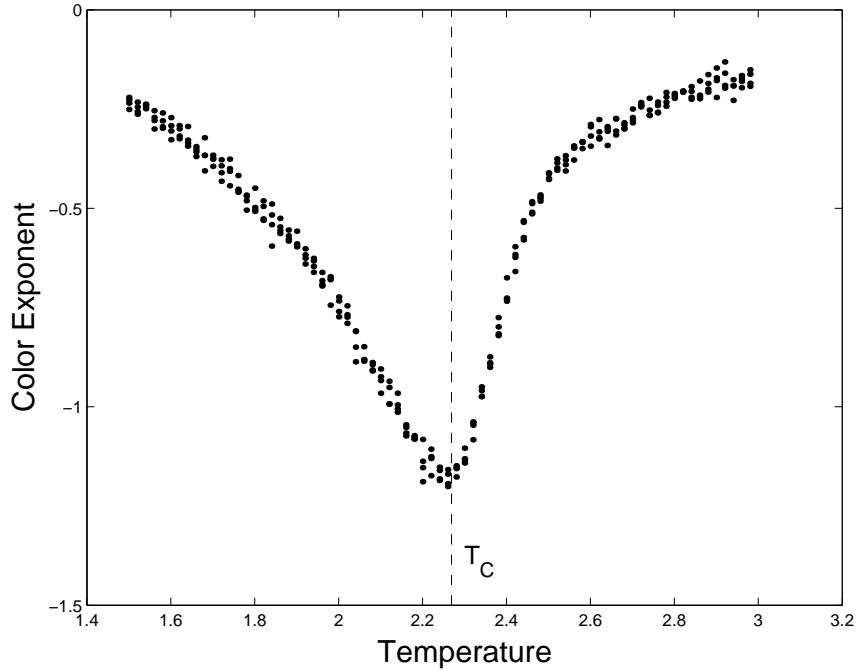,height=9cm}
  \caption{Color Exponents from Entropy Spectra, Ising system,
  four distinct i.c. at each temperature.}\label{Fin12}
\end{figure}
  This figure does not exhibit any burst of instability
   around the critical temperature $T_c = 2.27 J$,
   where a phase transition characterized by a fractal structure
   of clusters occurs \cite{parti2}. Moreover, the minimum is higher
  ($-1.2$ vs. $-1.6$).
  The qualitative difference between
  cluster geometry is therefore well reflected by Figs.11-12.
  More precisely, in the present case,
  the slow dynamics of small deformation of stripes ($\alpha \simeq -1.5$) is followed
  by a  substantial acceleration when domains start to merge. At $T_1$, temperature
  marking the appearance of very unstable stripes, dynamics slows down again.
  This pattern is confirmed also by Fig.4, by identifying slow/fast dynamics
  with long/short memory of correlations. Above $T_1$, the
   growth toward white noise is continuous: it does not give any particular relevance to $T_2$,
  possibly apart from a pseudo-fractality of very different nature with
  respect to the fractality found in NN Ising model. The exponents
  to compare are indeed $\alpha \simeq -1.2$ (Ising) vs.
  $\alpha\simeq -0.4$ (PFD). We stress again that this
  comparison is not between exponents at comparable temperatures,
  i.e. at the transition, but between the ``more fractal
  patterns'' in the two models (effective in a case, virtual in the other).

Some additional considerations on the thermal aspects of previous
results could be useful. As it is well known, the specific heat
may be calculated as
\begin{equation}\label{spheat}
C_V = \frac{<H^2>- <H>^2}{Nk_BT^2}~,
\end{equation}
$k_B$ being the Boltzmann constant (here $k_B=~1$). Labelling  the
exchange and the dipolar contributes in the Hamiltonian
(\ref{hamilton}) by indexes $e$ and $d$ respectively, we write:
\begin{equation}\label{sepheat}
C_V= \frac{<H_e^2>-<H_e>^2}{Nk_BT^2} +
\frac{<H_d^2>-<H_d>^2}{Nk_BT^2} + 2~\frac{<H_eH_d>-<H_e> <H_d>
}{Nk_BT^2}~.
\end{equation}
The first two terms in the r.h.s., we denote $C_e$ and $C_d$, have
the form of a specific heat for the exchange and dipolar
Hamiltonian respectively (of course, they are not!). The last
term, say $C_{ed}$, is a sort of correlation. Since $C_V= C_e+ C_d
+ C_{ed}$, one may look for the origin of peaks at $T_1$ and $T_2$
by observing separately $C_e+C_d$ and $C_{ed}$. Actually, both
these quantities give neat evidence of a peak at $T_1$, and no
evidence at all of a peak at $T_2$, as shown in Fig.13,
\begin{figure}[htbp]
  \centering
  \epsfig{file=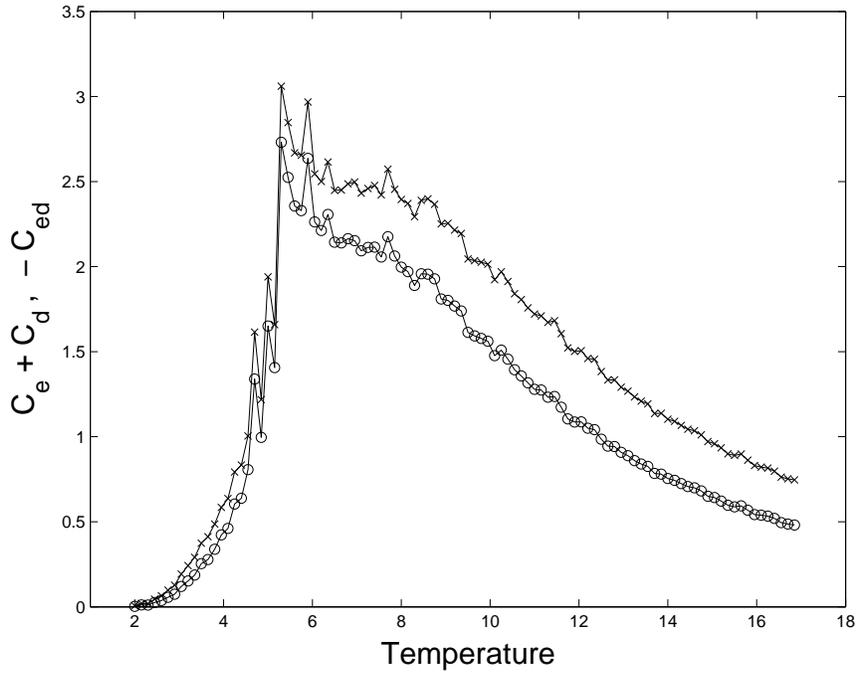,height=9cm}
  \caption{Contributes to the specific heat: $C_e+C_d$ (crosses), $-C_{ed}$ (circles),
   mean values over four i.c..}\label{Fin13}
\end{figure}
where  a close correlation between $C_e+C_d$ and $C_{ed}$ appears
in the whole range.
\begin{figure}[htbp]
  \centering
  \epsfig{file=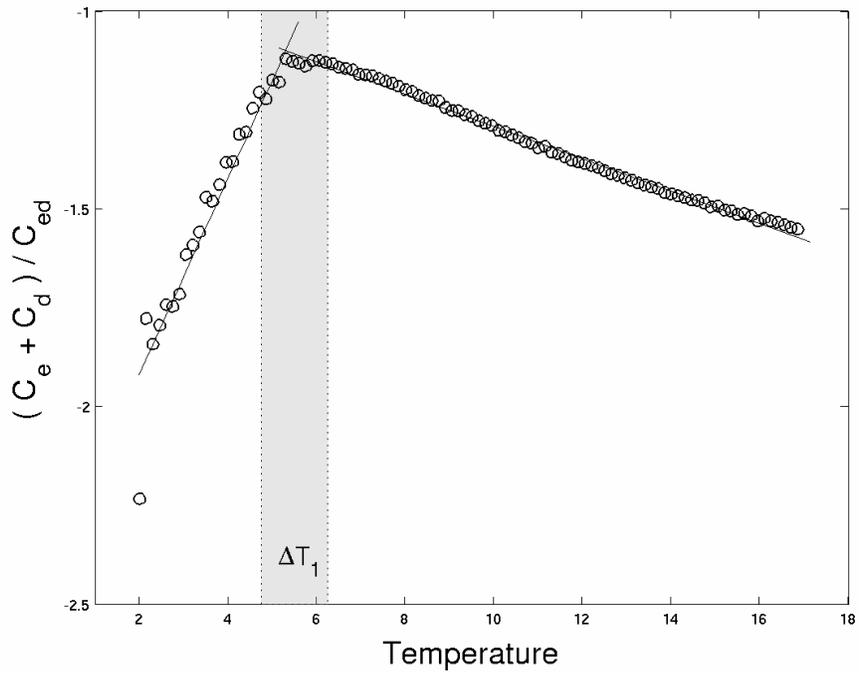,height=9cm}
  \caption{Ratio  $(C_e+ C_d)/C_{ed}$, same parameters
  as in Fig.13.}\label{Fin14}
\end{figure}
Moreover, the ratio $(C_e+C_d)/C_{ed}$, given in Fig.14, indicates
quite clearly that there exist two distinct regimes of
proportionality, $T_1$ being once again the turning point of them.

All conspire in saying that $T_2$ has no thermodynamic relevance.
This conclusion completely agrees with the geometrical
characterization suggested by entropy and distances: the maximum
of $C_V$ at $T_2$ seems to indicate a balancing point between
growing and decreasing contributes related to the smooth
fragmentation process of clusters.\\

 \vskip 30.0 pt

\noindent{ \bf 5. Previous results revisited }

 We recall that previous authors, in order to
evaluate the orientational symmetry of the striped states,
introduced two domain order parameters, $O$ and $\eta$.

In the dual lattice, let $n_h$ and $n_v$ be  the number of
horizontal and vertical sides along the cluster boundaries
($n=n_h+n_v$ is therefore the total border length, or total
perimeter).  The  first parameter $O$, introduced in \cite{booth},
is the  time averaged  difference
\begin{equation}\label{DOP}
  O = ~<{{n_h - n_v}\over {n}}>
\end{equation}
It estimates the deviation from an isotropic distribution of sides
in the clusters. In the purely striped domain, there are only
horizontal or vertical sides, so that $O = \pm 1$ (depending on
the initial orientation), while the parameter must be $0$ in an
isotropic configuration. Isotropy is expected to hold not only in
the disordered phase at high temperature, but just after the
stripes breakdown. What one actually sees in Fig.15, starting e.g.
with $O = 1$ for $T <2$, is that the parameter weakly decreases up
to temperatures where a preferential orientation clearly persists.
But the subsequent transition to $0$, just around $T_1$, is not
smooth at all, presenting a remarkable oscillation of sign, as if
the remainder of the stripes suddenly changed orientation for long
time intervals. For all $L$, the sign oscillation interval
coincides with the peak interval $\Delta T_1$ of the specific
heat. For both $C_V$ and $O$, this interval is expected to narrow
in the thermodynamic limit.
\begin{figure}[htbp]
\begin{center} \epsfig{file=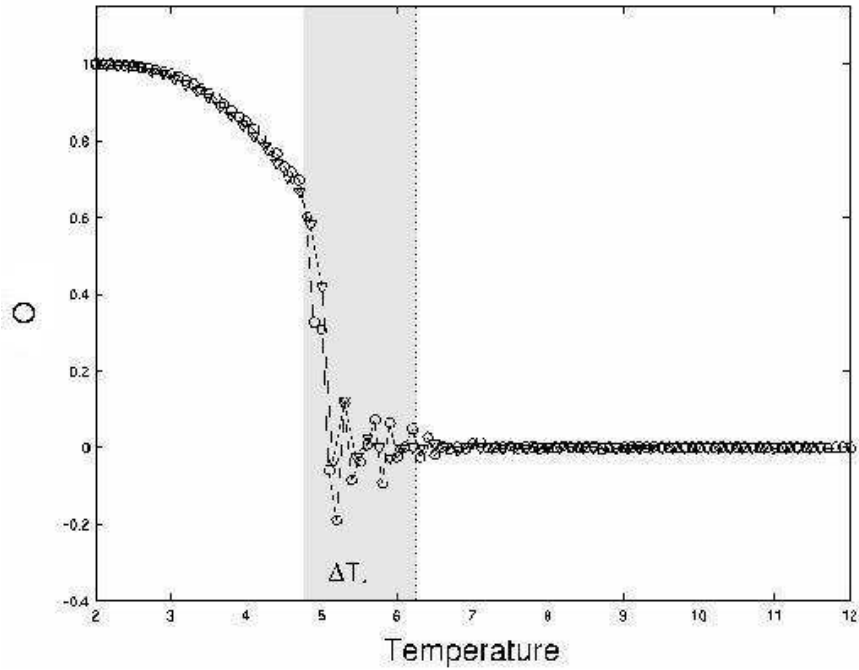,height=9cm} \caption{ Parameter $O$,
or first domain order parameter, for $L=32$ (circles) and $L=64$
(triangles),  mean values over four i.c..}\label{Fin15}
\end{center}
\end{figure}
 Except for these fluctuations around the value zero in the critical
region, Fig.15 recovers Fig.4 of \cite{booth}, apparently built up
with the absolute values of the parameter.  Even if such
inversions were a finite size effect, as suggested by the
amplitude of oscillations decreasing with $L$, this attitude is a
remarkable signature of the way the stripes collapse in the
transition region. This oscillatory phenomenon could be analogous
to the magnetization inversion observed in the NN Ising model, for
finite size lattices, when fractal patterns begin to take place
approaching the transition temperature. Note that in PFD this
orientation incertitude takes place mostly  after $T_1$, in a
range of temperatures where stripes are still melting into two big
clusters with small fragments, as shown by the entropy.

We recovered also the parameter $\eta$. Assuming $s_{i,j} \equiv
s_{\kk}$ the spin variable, $\eta$ is defined by formula (2) in
\cite{ifti}, i.e.:
\begin{equation}\nonumber
\eta = \frac{1}{N} \langle | \sum_{j}|\sum_{i}s_{i,j}|-
\sum_{i}|\sum_{j}s_{i,j}||\rangle~.
\end{equation}
 Also this parameter is proven to be effective in detecting the
first temperature (see. Fig.3 in \cite{ifti}), but it cannot
capture the oscillatory phenomenon revealed by $O$.

\vskip 15.0 pt

\noindent{ \bf 6. Conclusions}

The observation of cluster dynamics and the related statistical
properties  give clear evidence to the following points:
\begin{itemize}
  \item[1.] the peak for $C_V$ at $T_1$ is correlated to a melting
  process of stripes without any occurrence of fractality, a
  behavior quite different with respect to the NN Ising model;
  \item[2.] some analogy between the two models seems to exist in
  the attitude to sudden inversion of stripe orientation or
  magnetization respectively, as finite size effects. Such an
  analogy does not imply any similitude in the cluster geometry:
  in one case, the orientation anisotropy keeps a
  memory of the stripped structure, while in the other case
  the inversion of magnetization  has
  to do with the onset of fractality at $T_c$;

  \item[3.] colored noise is present in both models at their
  transitions (Figs.11-12), in a quite different fashion:
  in the PFD  case, with a wide range of values
  as a consequence of intermittent melting phenomena with small
  perturbations at boundaries (see e.g. Fig.3),
  making spectra at $T_1$ still dependent on i.c. (this intermittency
  may be read also in the entropy standard deviation); in Ising,
  as a counterpart of fractal dynamics, almost
  independent of i.c. at $T_c$;

  \item[4.] standard deviations for Rohlin and Hamming distances
  are different too in PFD and Ising systems. However, in both models, the Hamming SD
  have the same scaling behavior of the total perimeter;

  \item[5.] as to the second peak shown in the $C_V$ diagram, only
  the SD of the Rohlin distance presents some peculiar behavior in
  the neighbors of $T_2$. It seems that dynamical regime at such
  temperatures corresponds to the fragmentation of macro clusters.
  Only here there is a possibility for fractal configurations,
  even if not detectable at the values of $L$ accessible to computations.
\end{itemize}

In conclusion, we have a good evidence from geometry that the
Ising ferromagnetic transition at $T_c$ and the PFD transition at
$T_1$ are of different nature. As to $T_2$, on the basis of our
geometric and dynamical indicators it cannot be recognized as a
transition temperature at all. Consistently with expectations, we
cannot therefore extend to our case the conjecture proposed in
\cite{macisaac} on the irrelevance of the long range interactions
for the Ising antiferromagnet class of universality.

\vskip 20.0 pt
 \noi {\bf Acknowledgments}

\noi We thank A. Tassi (Parma) for very useful discussions on the
subject. 

\newpage

 \noindent{\textbf{Appendix}}

 Let $\mathbf{M}$ be a $L\times L$
square lattice, where knots $({i,j})$ assume values in an alphabet
$\mathbf{K}$. A state or configuration on $\mathbf{M}$ is a whole
set ${\mathbf{a}} =\{a_{i,j}\}, a_{i,j} \in \mathbf{K}$. It is an
element of ${\mathcal{C}}= {\mathcal{C}(\mathbf{M})}$, the set of
all $| {\mathbf{K}}|^{L\times L}$ possible states of the lattice.
For instance, ${\mathbf{K}} = \{0,1\}$, fits the description of
Ising-like systems. The dual lattice (we shall equally denote
$\mathbf{M})$ is a $L\times L$ set of square cells corresponding
to the knots.

When the alphabet $\mathbf{K}$  itself is a metric space (e.g. a
numerical set with the usual $~|~x-y~|~$ distance),  one can
consider  in ${\mathcal{C}(\mathbf{M})}$ the Hamming distance
$d_H$ which, for configurations $\mathbf{a}$ and $\mathbf{b}$, is
defined by the functional
\begin{equation}\label{hamming}
d_H ({\mathbf{a}},{\mathbf{b}}) =   \sum_{i,j}
|~b_{i,j}-a_{i,j}~|~.
\end{equation}
We stress that the Hamming distance is sensitive only to actual
values of corresponding knots, not to their distribution or
neighborhood.

A \textit{path}, is a sequence of ``near'' knots, equivalent to a
sequence of cells having common sides in the dual description. A
\textit{connected} cluster is a set of knots with the same value
in $\mathbf{K}$ which are connected by a path. In the dual
lattice, clusters are connected but not necessarily simply
connected sets, made up of square cells. Since every cell belongs
to a single cluster, clusters $A_{k}$ are disjoint subsets of
$\mathbf{M}$ and $ \bigcup_{k} A_{k} = \mathbf{M}$. In other
terms, the clusters collection is a ``finite partition'' of
$\mathbf{M}$. The subsets $\{A_k\}$ of a partition are often
referred to as its ``atoms''. Let $\mathcal{Z(\mathbf{M})}$ denote
the set of all finite partitions of $\mathbf{M}$. The
correspondence $\Phi : \mathcal{C} \rightarrow \mathcal{Z}$
between a configuration $\mathbf{a} \in \mathcal{C}$ and the
clusters partition $\alpha\equiv (A_1,...,A_N) \in \mathcal{Z}$,
i.e. $\alpha = \Phi(\mathbf{a})$, is ``many to one'', since
configurations generated by permutations in $\mathbf{K}$ are
mapped into the same partition. If the cardinality of the alphabet
is $| {\mathbf{K}}| \geq 4$, because of Euler's Four Colour
Theorem for every partition $\alpha \in \mathcal{Z}$ there exist
$\mathbf{a} \in \mathcal{C}$ with  $\alpha =\Phi(\mathbf{a})$.
This is not true for the case ${\mathbf{K}} = \{0,1\}$ considered
in the present work.

 A probability measure $\mu$ may be introduced in
the algebra $\mathcal{M}$ of subsets of $\mathbf{M}$: for every $A
\in \mathcal{M}~$, $~\mu(A) $ is the normalized number of knots in
$A$. Standard operations on partitions may be recovered in
classical textbooks such as \cite{bill}\cite{AA}\cite{sinai}, or,
for our demands, in \cite{parti1}\cite{soc}. Here we only recall
the definition of Shannon entropy and Rohlin distance: Let
$\alpha=(A_1,...,A_N)$ be a partition: its Shannon entropy $H
(\alpha)$ is
\begin{equation}
\label{shannon}
  H(\alpha)= -\sum_{i=1}^N \mu(A_i)\ln \mu(A_i) ~.
\end{equation}
Note that the Shannon entropy depends only on the cluster
measures, not on their shapes. Shapes are taken into account by
\textit{conditional} entropy: if f $\beta=(B_1,...,B_M)$ is
another partition, the conditional entropy of $\alpha$ with
respect to $\beta$ is
\begin{equation}
  H(\alpha|\beta) = -\sum_{i=1}^N\sum_{k=1}^M \mu(A_i\cap B_k)\ln\frac{\mu(A_i\cap
  B_k)}{\mu(B_k)}~,
\end{equation}
 and the Rohlin distance $d_R$ is
\begin{equation}
\label{rohlin} d_R (\alpha,\beta)
=H(\alpha|\beta)+H(\beta|\alpha)~.
\end{equation}
This way, $\mathcal{Z}(\mathbf{M})$ is a metric space. The Rohlin
distance between two finite partitions expresses how different
they are. We also recall that there exists a method, called
``reduction process '', to amplify as far as possible the
non-similarity between partitions. This method is reminiscent of
cancellation of common factors between integers, justifying the
concept of ``rational partitions'' introduced in this context
(\cite{parti1}\cite{parti2}). However, for the model studied in
the present work, the reduction process proves to be unimportant,
and we shall disregard on it.

Hamming and Rohlin distances are not directly comparable. We
stress that the Hamming distance is between
\textit{configurations} and it is sensitive only to actual values
of corresponding knots, not to their distribution or neighborhood,
 whereas the Rohlin distance is between
\textit{partitions}, and therefore is sensitive to the cluster
shapes. In principle, $d_R$ and $d_H$ may give very different
information. With the binary $\{0,1\}$ alphabet, for instance,
complementary configurations have maximal Hamming distance ($d_H =
N$), while the corresponding partitions coincide ($d_R=0$).

 If a configuration $a \in \mathcal{C}$ has discrete evolution
 $~a, ~Ta, ~T^2 a,...~$, one can speak of ``configurations orbit''.
 The corresponding dynamics $\hat{T}$ on $ \mathcal{Z}$ is defined by
\begin{equation}
  \hat{T}\alpha=\hat{T}~ \Phi(a)= \Phi~(Ta)
\end{equation}
so that to a configurations orbit there corresponds a partitions
orbit. Clearly, the probability measure $\mu$ in $\mathcal{M} $ is
not preserved by the evolution, in the sense that clusters or
atoms are redefined at every step by the pointwise evolution, and
do not evolve in themselves. However, we are not interested here
in such indicators as Kolmogorov-Sinai entropy or Lyapunov
exponents, requiring a preserved measure. Observables $F$ are
defined at each time in $\mathcal{C}(\mathbf{M})$ or
$\mathcal{Z}(\mathbf{M})$, and they give rise to ``time series''
$\{x_k \} = \{F(T^k a) \}$ or $\{x_k\} = \{F(\hat{T}^k \alpha)
\}$. Such time series are the main objects of our investigations.
Typically, we shall consider
\begin{itemize}
  \item $x_k= H(\hat{T}^k(\alpha))$, i.e. the entropy time  series;
  \item $x_k=d_R(\hat{T}^k(\alpha),\hat{T}^{k-1}(\alpha) )$, i.e.
  the Rohlin distance time series;
  \item $x_k=d_H(T^k(a),T^{k-1}(a) )$, i.e. the Hamming distance
  time series.
\end{itemize}
This formalism applies in principle to every kind of lattice and
discrete dynamics, and could be easily extended to graphs.
However, a computational obstacle consists in the necessity of
handling the cluster borders, a difficult task for large lattice
sizes (and even more in dimension $d > 2$).

\end{document}